\documentclass[aps,prl,amsmath,showpacs,preprintnumbers,twocolumn,amssymb,groupedaddress]{revtex4-1}

\usepackage{graphicx}
\usepackage{mathptmx, textcomp}
\usepackage[latin1]{inputenc}
\usepackage{tabularx}
\usepackage{siunitx}
\usepackage{color}
\definecolor{bl}{rgb}{0, .1, .6}
\definecolor{rd}{rgb}{1,0,.2}

\usepackage[colorlinks=true, citecolor = bl, linkcolor = bl, urlcolor=bl, pdfborder={0 0 0}]{hyperref}
\usepackage{natbib}
\usepackage{tabularx}
\usepackage{lipsum}
\newcommand{\mum}{\SI{}{\micro\meter}}
\newcommand{\mus}{\SI{}{\micro\second}}
\newcommand{\nm}{\SI{}{\nano\meter}}

\newcommand{\Hz}{{\rm Hz}}
\newcommand{\g}{{\rm G}}

\newcommand{\ms}{{\rm ms}}
\newcommand{\be}{\begin{eqnarray}}
\newcommand{\ee}{\end{eqnarray}}
\newcommand{\abg}{a_{\rm bg}}

\newcommand{\add}{a_{\rm dd}}
\newcommand{\edd}{\epsilon_{\rm dd}}

\newcommand{\dy}{\textsuperscript{164}Dy}

\begin{document}

\title{
%Observation of quantum-stabilized solitons in a strongly dipolar Bose gas \\or\\
Observation of quantum droplets in a strongly dipolar Bose gas
%\\or\\
%Quantum fluctuations stabilized droplets of a strongly dipolar Bose gas\\
%or\\
%Observation of quantum fluctuations-stabilized droplets
}

\author{Igor Ferrier-Barbut}
\author{Holger Kadau}
\author{ Matthias Schmitt}
\author{Matthias Wenzel}
\author{Tilman Pfau}
\affiliation{5. Physikalisches Institut and Center for Integrated Quantum Science and Technology,
Universit\"at Stuttgart, Pfaffenwaldring 57, 70550 Stuttgart, Germany}

%\author{Pep Guardiola}
%\altaffiliation{These authors contributed equally to this work.}
%\affiliation{5. Physikalisches Institut and Center for Integrated Quantum Science and Technology,
%Universit\"at Stuttgart, Pfaffenwaldring 57, 70550 Stuttgart, Germany}
%

%\date{\today}

%\pacs{Romana}
\begin{abstract}
Quantum fluctuations are the origin of genuine quantum many-body effects, and can be neglected in classical mean-field phenomena. Here we report on the observation of stable quantum droplets containing $\sim$ 800 atoms which are expected to collapse at the mean-field level due to the essentially attractive interaction. By systematic measurements on individual droplets we demonstrate quantitatively that quantum fluctuations mechanically stabilize them against the mean-field collapse. We observe in addition interference of several droplets indicating that this stable many-body state is phase coherent.
\end{abstract}

\maketitle

Uncertainties and fluctuations around mean values are one of the key consequences of quantum mechanics. At the many-body level, they induce corrections to mean-field theory results, altering the many-body state, from a classical factorizable to an entangled state. Owing to their versatility, ultracold atom experiments offer numerous examples of interesting many-body states \cite{Bloch:2008}. Among these systems, bosonic superfluids are well studied. They are described in the weakly interacting regime by a mean-field energy density proportional to the square of the particle density $n^2$, with a negative prefactor in the attractive case. Since the seminal work of Lee, Huang and Yang \cite{Lee:1957}, it is known that interactions lead to a repulsive correction $\propto n^{5/2}$ owing to quantum fluctuations. Therefore an equilibrium between these two contributions can in principle stabilize an attractive Bose gas \cite{*[{}] [{ section 3.3, p. 29.}] Volovik:2009}. A similar stabilization mechanism using quantum fluctuations was proposed for an attractive Bose-Bose mixture in ref.~\cite{Petrov:2015}, which leads to the formation of droplets. In this reference liquid-like droplets are defined as the result of a competition between an attractive $n^{2}$ and a repulsive $n^{2+\alpha}$ term in the energy functional. Besides liquid helium droplets \cite{Dalfovo:1995}, such functionals are also used to describe atomic nuclei \cite{Bender:2003}. Here we study a strongly dipolar Bose gas where the attractive mean-field interaction is due to the dipole-dipole interaction (DDI). This system is known to be unstable in the mean-field approximation \cite{Komineas:2007}. We however show here that beyond mean-field effects lead to the stabilization of droplets.\\
\indent Our investigations are aimed at probing strongly dipolar Bose gases of \dy, that are characterized by a dipolar length $\add=\mu_0\mu^2m/12\pi\hbar^2\simeq131\,a_0$ where $a_0$ is the Bohr radius, with $\mu = 9.93\,\mu_B$ Dy's magnetic dipole moment, in units of the Bohr magneton $\mu_B$, $\hbar$ the reduced Planck constant and $m$ the atomic mass. The additional short-range interaction of \dy, characterized by the scattering length $a$ has been the focus of several papers \cite{Baumann:2014,Maier:2015a, Tang:2015,Maier:2015b}, and the background scattering length was measured to be $\abg=92(8)\,a_0$, modulated by many Feshbach resonances. Thus, away from Feshbach resonances at the mean-field level the dipolar interaction dominates with $\varepsilon_{\rm dd, bg}=\add/\abg\simeq1.45$. In a previous work \cite{Kadau:2016}, we have reported the observation of an instability of a dipolar BEC, the resulting state of this instablity is characterized by the existence of apparent droplets. These droplets cannot be explained by a stabilization by one-body quantum pressure \cite{Pedri:2005}, and as such are not solitons in the strict sense.\par
\begin{figure}
\begin{centering}
\begin{tabularx}{\columnwidth}{m{0.43\columnwidth}m{0.55\columnwidth}}
\includegraphics[width=.43\columnwidth]{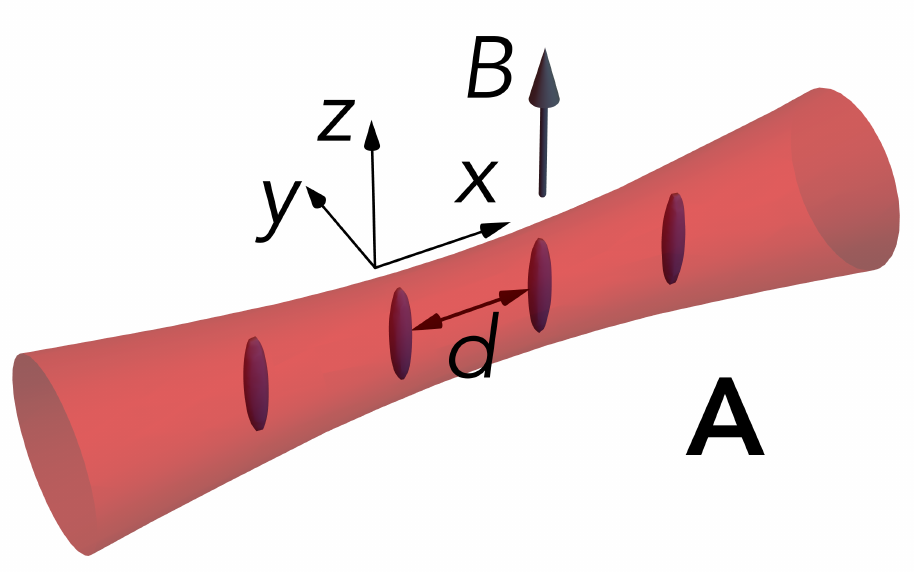}
  & \includegraphics[width=.55\columnwidth]{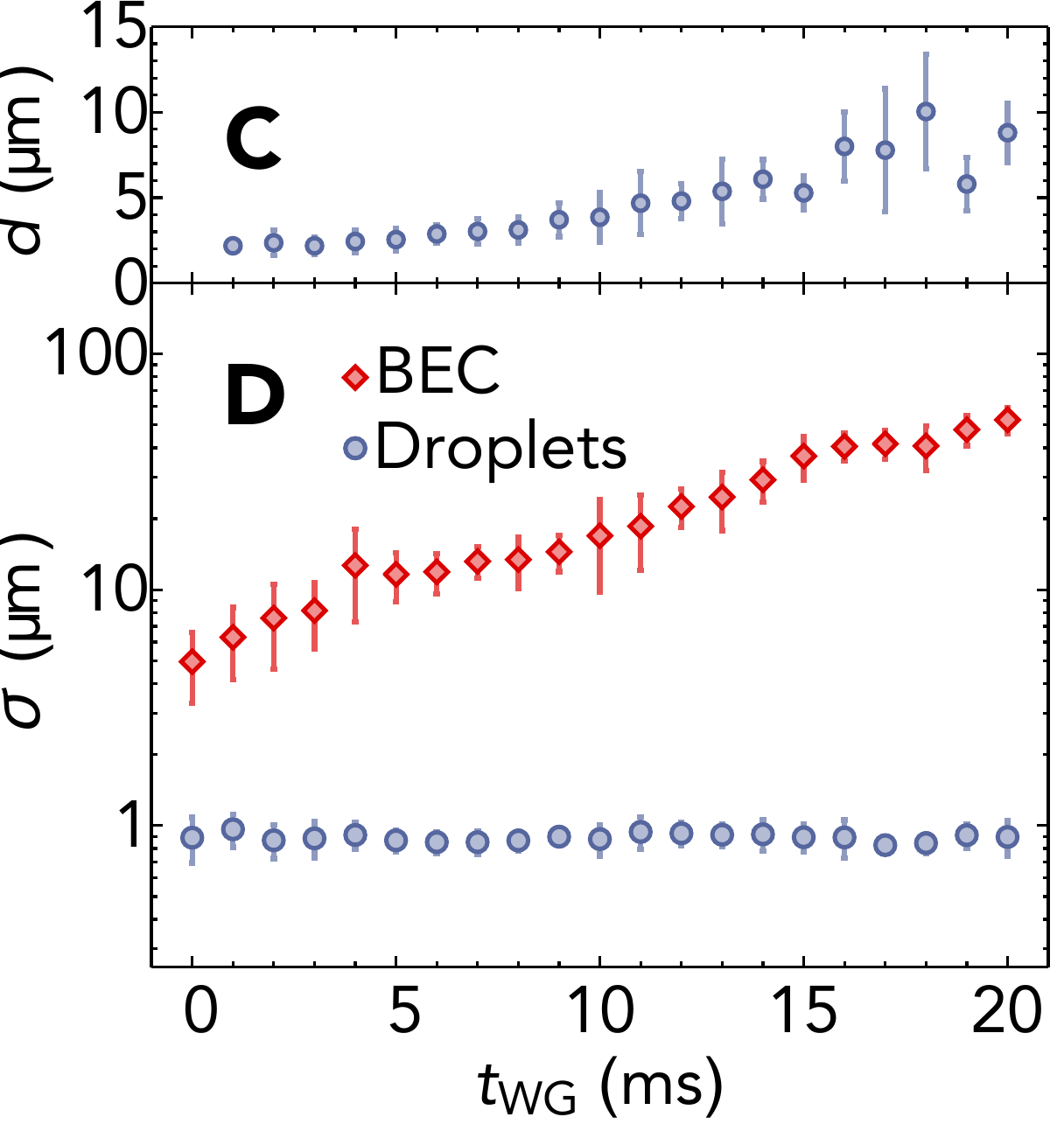}
\end{tabularx}
\includegraphics[width=\columnwidth]{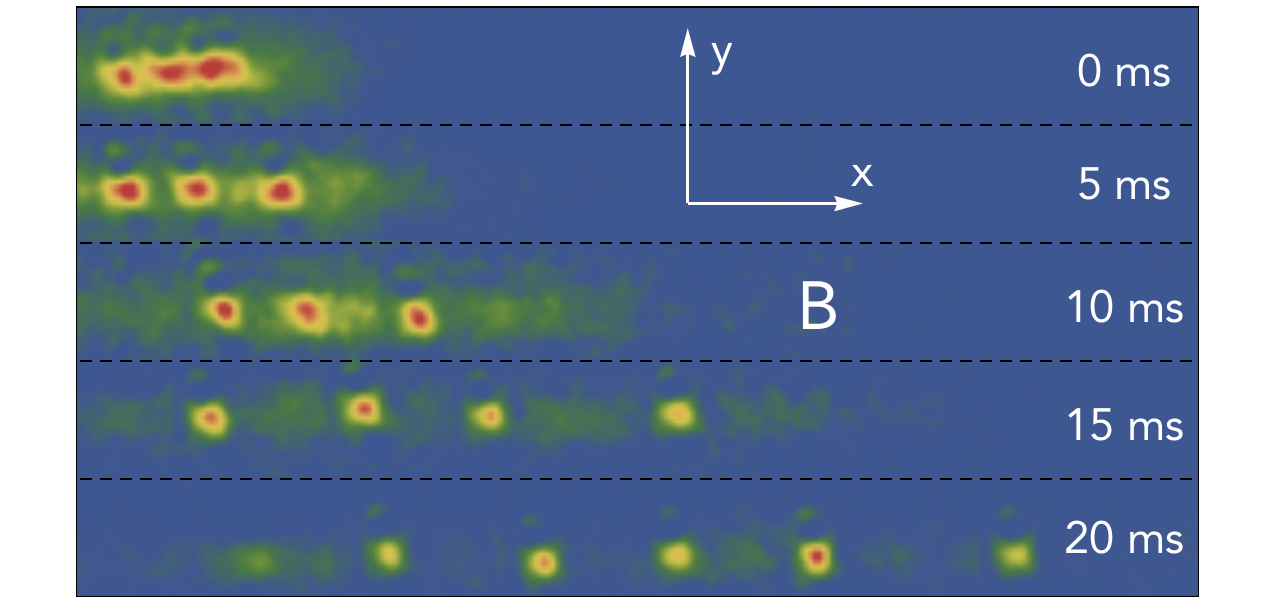}

\end{centering}
\caption{Quantum droplets of a dipolar Bose gas in a waveguide. A: Schematic representation of the droplets in the waveguide, the elongation along $z$ is represented, their separation $d$ is indicated. B: Examples of in-situ optical density (OD) images after release in the waveguide at the magnetic field $B_1 = 6.656(10)\,\g$. Images taken at times $t_{\rm WG} = 0,\,5,\,10,\,15,\,20\,\ms$ (top to bottom). The OD is normalized to the maximal OD in each image to improve visibility. C: Evolution of the mean separation $d$ between the droplets as a function of time. D: Evolution of the width $\sigma$ obtained from a gaussian fit to their density profiles (average of transverse and axial radii), blue circles. Compared with the evolution of the size of a BEC, red diamonds. The data in panels C and D is obtained by averaging at least 4 experimental realizations, error bars indicate statistical standard deviation The convention for the axes used through the paper is indicated in A.}
\label{Fig:SolitonsSize}
\end{figure}

\indent Here we isolate these droplets to unravel their nature. To perform our study systematically, we place them in a waveguide. This relaxes their confinement in one direction (along $x$) and thus supresses the effect of dipolar repulsion between the droplets. The waveguide is a single optical dipole trap that creates a tight confinement around the $x$-axis with frequencies $\nu_y = 123(5)\,\Hz$, $\nu_z=100(10)\,\Hz$. The release in this waveguide is performed in the following way (details of ramping procedures can be found in \cite{SupMat}): We create a BEC containing $\sim10\times10^3$ atoms in a crossed optical dipole trap at a magnetic field along the vertical ($z$) axis $B_{\rm BEC} = 6.962(10)\,\g$, we then lower the field to $B_1 = 6.656(10)\,\g$ in $1\,\ms$, from which a wait time of $15\,\ms$ follows. At $B=B_1$ ($B=B_{\rm BEC}$) using $\abg=92\,a_0$ and our knowledge of the Feshbach resonances \cite{SupMat}, we get $a=95(13)\,a_0$ ($a=115(20)\,a_0$). Then, one dipole trap is turned off and the other one ramped-up to higher power in $1\,\ms$. The trap has a too weak confinement to hold the atoms in the $x$ direction and the cloud starts moving. We then image it as function of time in the waveguide $t_{\rm WG}$ using high-resolution ($1\,\mum$) imaging. We observe the following, illustrated in figure \ref{Fig:SolitonsSize}: First the condensed fraction remains fragmented into up to six droplets and down to one droplet. Some atoms originally in the BEC do not form droplets, this fraction of atoms is hard to quantify since it is hard to tell apart from a thermal fraction in our images. Second, during the evolution time the initial confinement energy is turned into relative kinetic energy and these droplets move away from each other. We observe an in-situ size limited by our resolution (gaussian width $\sigma \simeq 900\,\nm$ roughly identical in the $x$ and $y$ directions), which does not evolve during $20\,\ms$. If we perform the same sequence but keeping the field at $B_{\rm BEC}$, we observe that the BEC does not separate into droplets and expands as a whole in the waveguide, in $20\,\ms$ its axial size is increased by a factor 10 (fig.\,\ref{Fig:SolitonsSize}\,D red diamonds).
%\tocorrect{In the same time the axial size of a BEC (obtained by keeping the magnetic field at the value $B_{\rm BEC}$) increases by a factor $10$ (fig.\,\ref{Fig:SolitonsSize}\,D red diamonds).}
At $B = B_1$, the number of atoms in the droplets is $N=800(200)$. The facts that a single droplet appears to be stable and, when there are several of them, their size does not significantly increase while their distance is multiplied by $4$, indicates that they are self-confining. Note that we also observe these droplets on the low-field side of a resonance at $B=1.2\,\g$.\\

Since the confinement is too weak in the long direction to observe droplets for longer times, we perform a second set of experiments keeping a very weak confinement in the $x$ direction ($\nu_x=14.5(1)\,\Hz$, see \cite{SupMat}), thus the trapping potential takes a prolate cigar shape still perpendicular to $\vec B$ with aspect ratio $\nu_{y,z}/\nu_x\simeq8$. We observe that in this trap, the droplets equilibrate at long times $t>100\,\ms$ at an average relative distance $d=2.5(5)\,\mum$, obtained from 10 experimental realizations.
Furthermore when we first adiabatically load a BEC in the prolate trap and then ramp from $B_{\rm BEC}$ to $B_1$, we observe the same distance. This distance is smaller than the length obtained by a simple analysis assuming point-like dipoles in a harmonic trap  $l_x = \left(3N\mu_0\mu^2/2\pi m\omega_x^2\right)^{1/5}\simeq4.5\,\mum$, indicating that the droplets cannot be considered as point-like. With a more refined analysis developed in \cite{SupMat} using a gaussian ansatz with radial symmetry around $z$ for the density distribution inside a droplet, we calculate the dipole-dipole repulsion. We thus obtain that a distance of $d=2.5(5)\,\mum$ is obtained for elongated droplets with $\sigma_z=2.5(5)\,\mum$ and radial size $\sigma_r\lesssim500\,\nm$. Finally we observe lifetimes of several hundreds of $\ms$, similar to what we reported in \cite{Kadau:2016} which confirm a strong stabilization mechanism.\par\indent
Given the strong elongation of the droplets along the $z$ direction, the dipolar interaction is mainly attractive and since $\edd=\add/a>1$ this attraction is stronger than the short-range repulsion, such that overall the interactions are mainly attractive. The droplets are thus expected to be unstable at the mean-field level \cite{Koch:2008}. We observe that first the gas locally collapses, before this collapse is arrested at high densities finally forming droplets. This means that the density dependence of the stabilizing mechanism is stronger than that of mean-field two-body interactions. Importantly, our present work shows that this mechanism is local and not due to any long-range effect between droplets. Two works have postulated the existence of a three-body conservative repulsion \cite{Xi:2016,Bisset:2015} with mean-field energy density $\propto n^3$.\\
\indent However these works neglect beyond mean-field effects. As stated above the energy density $e$ for these effects is $e\propto n^{5/2}$. This correction has been measured in contact-interacting Bose gases \cite{Papp:2008,Navon:2011}. Here we must take both contact repulsion and the DDI into account. Using the results of \cite{Scutzhold:2006, Lima:2011, Lima:2012} the beyond mean-field correction to the chemical potential $\mu=\frac{\partial e}{\partial n}$ for a dipolar gas is given by $\mu_{\rm bmf}\simeq\frac{32\,gn}{3\sqrt\pi}\sqrt{na^3}(1+\frac32\edd^2)$ where we have taken the lowest order expansion of the $Q_5$ function of  ref.\,\cite{Lima:2012} since $\edd$ is close to 1. Doing this we effectively neglect the imaginary part which is very small compared to the real part, such that a long lifetime is still ensured, though it is only in a metastable equilibrium. This beyond mean-field term is to be compared with the mean-field contact interaction contribution $\mu_{\rm c, mf}$ and the DDI one $\mu_{\rm dd, mf}$. Using a Thomas-Fermi approximation (which neglects kinetic energy) for a droplet with gaussian density distribution, the contribution at the center of the droplet is $\mu_{\rm c, mf}=gn_0$ for the contact interaction where $g=4\pi \hbar^2a/m$ and $n_0$ is the peak density. The dipolar interaction contribution is $\mu_{\rm dd, mf}=-gn_0\,\edd\,f_{\rm dip}(\kappa)$ \cite{Lahaye:2009} with $\kappa =\sigma_r/\sigma_z$, it thus depends on the elongation of the droplet along the field direction through the function $f_{\rm dip}(\kappa)$ which can be found in \cite{SupMat}. Using an aspect ratio equal to our experimental upper bound $\kappa=0.2$ one has $f_{\rm dip}(\kappa)=0.83$ such that the dipolar attraction dominates mean-field contributions for $\edd\geqslant1.2$ or $a\leqslant110\,a_0$ \footnote{Since $f_{\rm dip}(\kappa)$ is a monotonically decreasing function of $\kappa$, at lower aspect ratios than $0.2$, the mean-field dipolar contribution is larger and renders the droplets more unstable in the mean-field approximation.}.
%\tocorrect{At our aspect ratio this function is very close to one and since $\edd>1$, the dipolar attraction dominates mean-field contributions.} 
The mechanical stability condition is $\frac{\partial\mu}{\partial n}\geqslant0$. At center in the gaussian ansatz we get 

\begin{figure}[htbp]
\includegraphics[width=\columnwidth]{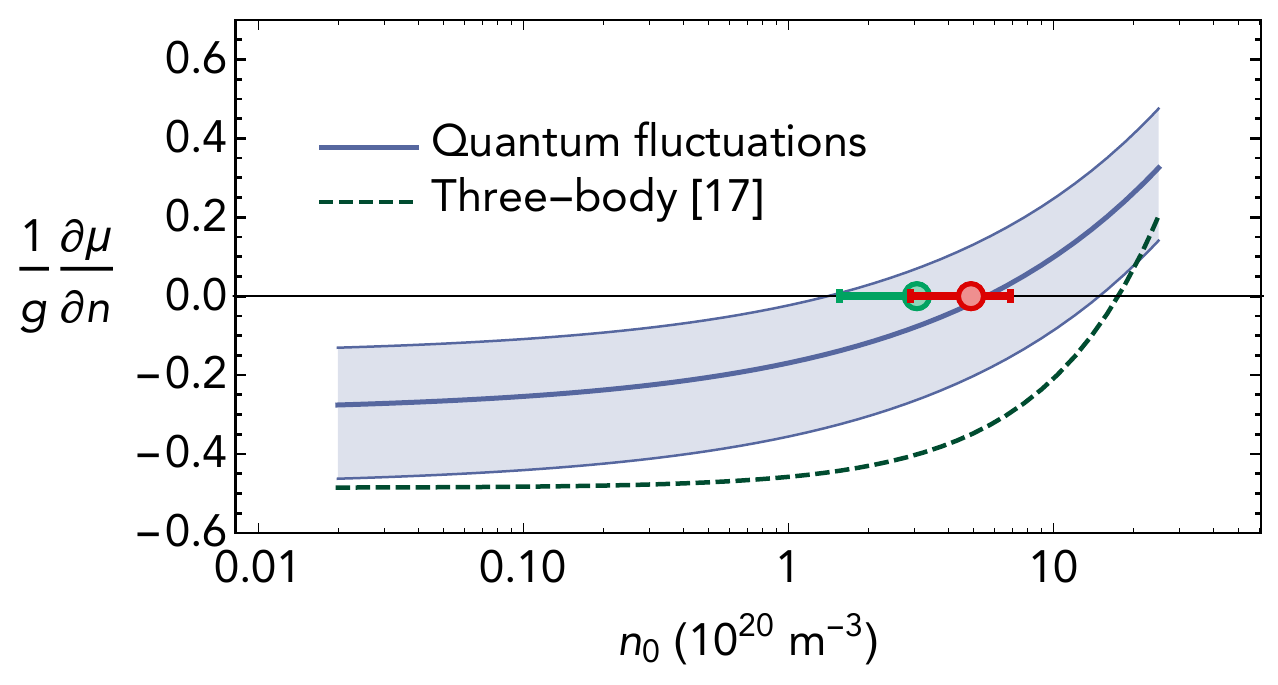}
\caption{Derivative of the chemical potential with respect to density as a function of density, at the center of a droplet using a gaussian ansatz, eq.~(\ref{Eq:compressibility}) ($g=4\pi\hbar^2 a/m$). The blue shaded region expresses our uncertainty on the scattering length. Negative values imply mechanical instability. The experimental value obtained from expansion measurements (fig.\,\ref{Fig:ToF}) is shown in red assuming a gaussian distribution and in green assuming an inverted parabola. The dashed line shows the same quantity obtained using a three-body repulsion using parameters from ref.\,\cite{Bisset:2015}, which stabilizes at a higher density.}
\label{Fig:Compressibility}
\end{figure}

\vspace{-.5cm}
\begin{small}
\be
\left.\frac{\partial\mu}{\partial n}\right|_{\boldsymbol r=0}=g\left(1-\edd\, f_{\rm dip}\left(\kappa\right)+16\sqrt{n_0a^3/\pi}\,(1+\frac32\edd^2)\right)\label{Eq:compressibility}
\ee
\end{small}
\normalsize
where $n_0$ is the peak density. Note that if one assumes an inverted-parabola density distribution, then one obtains the same result \cite{Odell:2004}. We plot this function in fig.\,\ref{Fig:Compressibility} using $a=95(13)\,a_0$ and $\kappa=1/10$ (this $\kappa$ value is a factor two below the experimental upper bound, it yields $f_{\rm dip}(\kappa)=0.94$). One can clearly see that since $\edd$ is close to one, though the attraction dominates, the two mean-field contribution nearly balance each other which leads to a major role for beyond mean-field effects, a very similar situation to the one considered in ref.~\cite{Petrov:2015}. From eq.~(\ref{Eq:compressibility}) one easily derives that the central density stabilizes at the value 
\be
n_0=\frac{\pi}{a^3}\left(\frac{\edd f_{\rm dip}(\kappa)-1}{16(1+3\edd^2/2)}\right)^2,\label{Eq:Centraldens}
\ee
thus in our approximation, stability is reached at densities $n_0\gtrsim10^{20}\,{\rm m}^{-3}$. Eq.\,(\ref{Eq:Centraldens}) is striking because the central density does not depend on atom number but only on $a$ and very weakly on $\kappa$ \footnote{At our aspect ratios $f_{\rm dip}(\kappa)$ depends only weakly on $\kappa$ as is visible in the Supplemental Material}, which is characteristic of a liquid-like state. Neglecting quantum fluctuations and assuming a three-body repulsion ($\mu_{\rm 3b,mf} = \hbar\kappa_3n^2/2$), this density becomes $n_0= g\frac{\edd f_{\rm dip}(\kappa)-1}{\hbar\kappa_3}$. Using parameters from \cite{Bisset:2015} ($a=82.6\,a_0$, $\kappa_3=5.87\times10^{-39}\,{\rm m}^{6}/{\rm s}$) we get $n_0=17\,\times10^{20}\,{\rm m}^{-3}$, in very good agreement with full simulations results \cite{Bisset:2015}, at these densities however beyond mean-field effects cannot be neglected. In addition such a high value for $\kappa_3$ is very hard to justify. It is very probable that $\kappa_3$, which is the real part of the three-body coupling constant, lies close to its imaginary part which is the three-body recombination constant $L_3$. Observing lifetime of BECs, we have an upper-bound $L_3\lesssim10^{-41}\,{\rm m}^{6}/{\rm s}$, which implies an experimentally irrelevant stabilizing density $n_0>10^{23}\,{\rm m}^{-3}$. Our experimental observations developed above imply a lower-bound on central density $n_0\geqslant10^{20}\,{\rm m}^{-3}$, given our imaging resolution, we cannot observe smaller droplet radii and higher densities. For a better estimate of the density, we turn to expansion experiments. \\

\indent The mechanisms at work in the droplets can indeed be further explored by observing their time-of flight expansion in free space. In principle pure liquid droplets in the absence of trapping should reach an equilibrium with an absence of growth \cite{Petrov:2015, Xi:2016}. On the other hand time-of-flight expansion under dipolar interaction is-non trivial but well studied \cite{Giovanazzi:2006,Lahaye:2007}, and it is modified by beyond mean-field effects \cite{Lima:2012}, these effects are isotropic and counteract magnetostriction. Mean-field hydrodynamic equations could not describe the expansion of our droplets. In our experiment, we perform such measurements by turning-off the waveguide trap after $4\,\ms$. In order to keep the atoms at the focal position of our imaging system, we apply a magnetic field gradient that compensates gravity, and image the atoms at various times after release, fig.\,\ref{Fig:ToF}\,A,\,B. We record thus the size in the $x$ and $y$ direction as a function of time. The sizes undergo a linear growth with rates $\dot \sigma_x=0.17(3) \,\mum/\ms$, $\dot \sigma_y=0.24(3)\,\mum/\ms$ \cite{SupMat}. We qualitatively express the expansion dynamics in terms of released energy $E_i=\frac12 m \dot \sigma_i^2$ \cite{Holland:1997}, we get $E_y=0.09(1)\,\hbar\omega_y$, $E_x=0.045(4)\,\hbar\omega_y$. Such energies are remarkably low which demonstrates that kinetic energy plays only a marginal role as expected, however a full theory is presently not available to describe the free-space dynamics after release.\\
\begin{figure}[htbp]
\includegraphics[width=1\columnwidth]{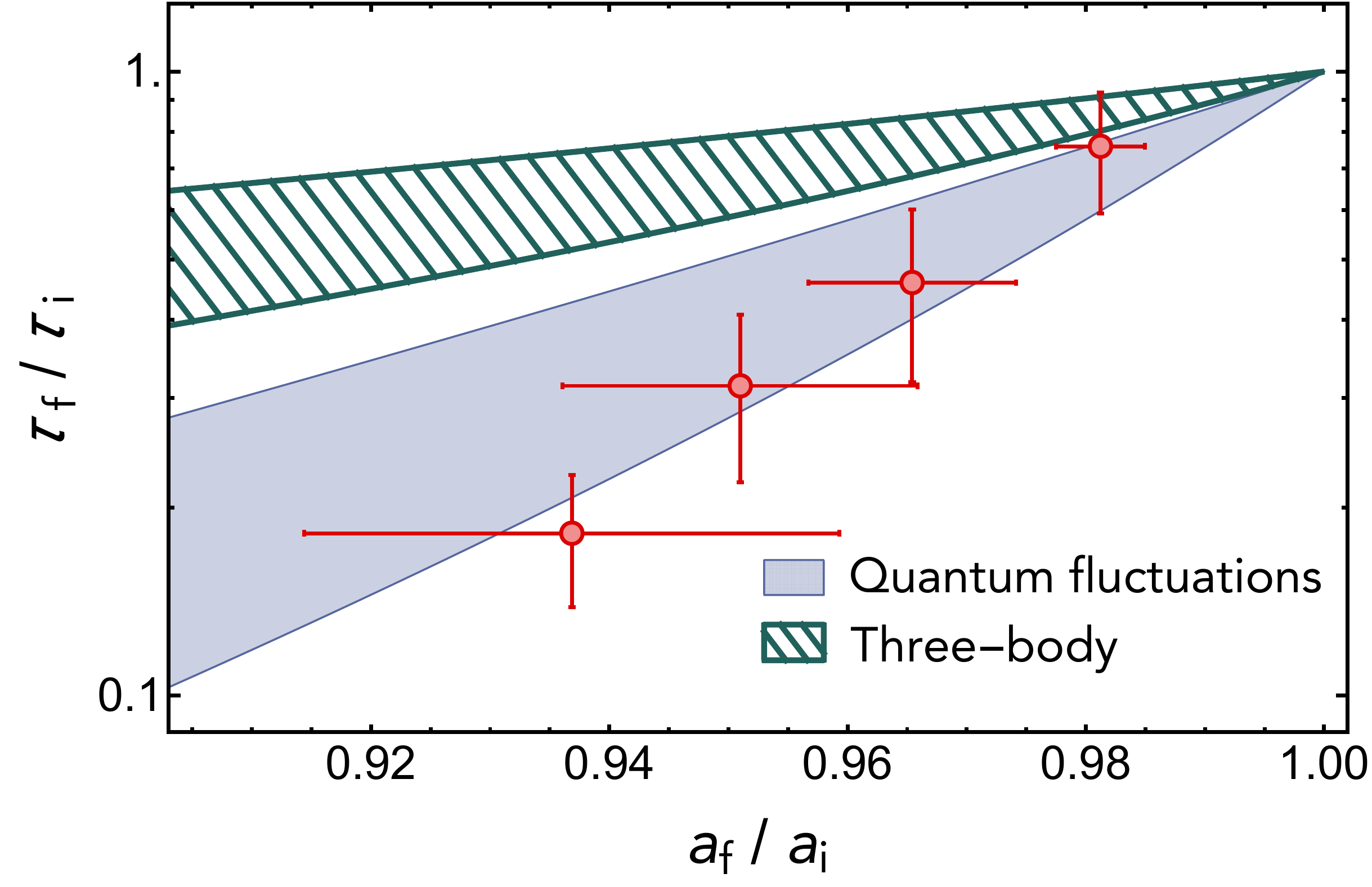}
\caption{Ratio of the lifetime $\tau_{\rm f}/\tau{\rm i}$ of the droplets between scattering lengths $a_{\rm f}$ and $a_{\rm i}$. We use here $a_{\rm i}=94(12)\,a_0$ obtained at $B_{\rm i}=6.573(5)\,\g$. The data points are taken down to $B_{\rm f}=6.159(5)\,\g$. The filled blue and green hatched areas represent the expected scaling using quantum fluctuations and three-body repulsion respectively, taking into account the uncertainty range on the droplets aspect ratio: $0\leqslant\kappa \leqslant0.2$.}
\label{Fig:UniversalScaling}
\end{figure}

\indent To circumvent the absence of a model for free-space dynamics, we perform a new set of experiments. It consists of the same procedure, but at the time of release, the magnetic field is quenched (in $50\,\mus$) from $B_1$ to a higher value $B_{\rm ToF}$ inducing a change in scattering length $\Delta a=a(B_{\rm ToF})-a(B_1)$, while the DDI remains unchanged. In this case, the expansion rate is strongly increased. Given the quench time, the initial density does not have time to adapt to the interaction quench. One thus expects that the change in released energy is given by $\Delta E\simeq\frac1N\,\int d\vec r\;\frac{\Delta g}{2}  n^2=\Delta g \,n_0/4\sqrt{2}$
where we have used again a gaussian ansatz, and $\Delta g= 4\pi\hbar^2\,\Delta a/m$. Since we are dealing with the difference in total energy here, the variation of the beyond mean-field corrections is negligible. Thus, since $\Delta a(B)$ is known, we can extract a value for $n_0$ from the observed change in $\Delta E$. From these measurements detailed in \cite{SupMat}, given our uncertainty on $a(B)$ we obtain $n_0=4.9(2.0)\times10^{20}\,{\rm m}^{-3}$. If instead of the gaussian ansatz we use an inverted parabola, then we get $\Delta E=2\Delta g \,n_0 /7$, from which we obtain $n_0=3.0(1.5)\times10^{20}\,{\rm m}^{-3}$. Both values are compatible with the lower-bound extracted from in-situ imaging, we represent them in fig.\,\ref{Fig:Compressibility}. The measured density is thus in agreement with the stabilizing density due to quantum fluctuations.\\
\indent However this does not probe the scaling behaviour of the density as a function of $a$. As evident in eq.\,(\ref{Eq:Centraldens}), this scaling is very strong. In turn, three-body recombination in the droplets scales very strongly with $a$, indeed, since the density does not depend on atom number, three-body losses lead to an exponential decay with lifetime $\tau=1/L_3\,\langle n^2\rangle$ \cite{SupMat}. In particular, $\tau$ decreases when $a$ decreases. To cancel uncertainties on $L_3$ and on the exact density distribution one simply needs to measure the ratio in lifetime $\tau_{\rm f}/\tau_{\rm i}$ between two different scattering lengths or magnetic fields $B_{\rm i}$ and $B_{\rm f}$, which, assuming constant $L_3$, is simply given by $\tau_{\rm f}/\tau_{\rm i}=\frac{\langle n_{\rm i}^2\rangle}{\langle n_{\rm f}^2\rangle}=\frac{n_{0, \rm i}^2}{n_{0, \rm f}^2}$. One can easily show that for fixed $\kappa$ this ratio is a function of only two parameters: $(\frac{a_{\rm f}}{a_{\rm i}},\frac{\add}{a_{\rm i}})$, in particular assuming 3-body repulsion, it is independent on $\kappa_3$, we give this function in \cite{SupMat}. Thus using a fixed $a_{\rm i}=94(12)\,a_0$ ($B_{\rm i}=6.573(5)\,\g$), in fig.\,\ref{Fig:UniversalScaling} we represent $\tau_{\rm f}/\tau_{\rm i}$ vs. $a_{\rm f}/a_{\rm i}$. This figure is striking, while we vary the scattering length by less than $10\%$, the lifetime is divided by a factor 5. Furthermore the data points are incompatible with the scaling predicted by three-body repulsion while without any fit parameter they follow the scaling predicted using quantum fluctuation within experimental uncertainties. The small deviation to lower lifetimes can be accounted for by a weak variation of $L_3$ \footnote{Measurements of $L_3$ will be presented in a future publication}. This demonstrates unambiguously that quantum fluctuations constitute the stabilizing mechanism. The conclusion we drew here is reinforced by numerical simulations reported shortly after the first submission of this paper in \cite{Wachtler:2016}. \\
\begin{figure}[htbp]
\includegraphics[width=1\columnwidth]{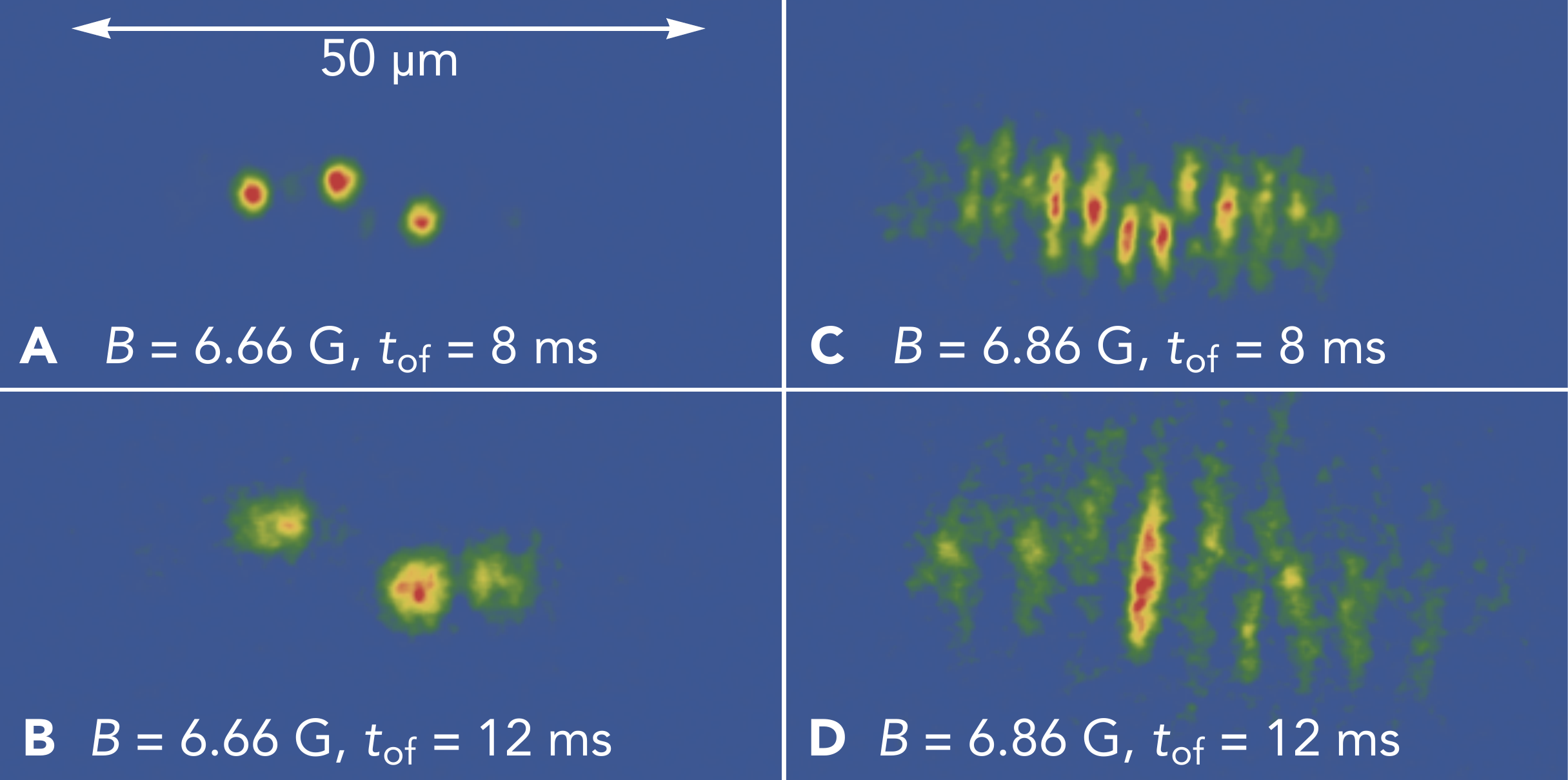}
\caption{Time of flight expansion measurements. The field is held at $B_1=6.656(10)\,\g$ until release when it is quenched to a different value. A,B Images where the field is kept at $B_1$ during expansion and C, D quenched to $6.86\,\g$. In A, B one sees expanding droplets, whereas  in C, D, they overlap and clear interference fringes appear along the $x$ axis while we can still measure the expansion size in the $y$ direction.}
\label{Fig:ToF}
\end{figure}
\indent Finally we observe that the droplets have internal phase coherence. Indeed for ``fast'' expansion dynamics obtained when quenching $B$ during time-of flight, the size of the expanding droplets becomes comparable to or larger than their relative distance so that neighbouring ones overlap. In this case we observe matter-wave interference fringes as exemplified in fig.\,\ref{Fig:ToF}\,C,\,D. The presence of these fringes demonstrates that each droplet individually is phase coherent and thus superfluid. Their observation opens the door to studies of the relative phase coherence between droplets. In the present case we do not observe fringe patterns that allow us to measure the droplets relative phase, but this is mainly due to shot-to-shot noise in the in-situ position and relative spacing of the droplets since we are not yet in the far-field regime.\\
\indent Future studies with fixed in-situ conditions prior to time-of-flight could bring insight in the phase coherence of an ensemble of droplets, even in the case of a high number of them \cite{Hadzibabic:2004}. Our measurements reported here have established the existence of a novel system forming droplets stabilized by quantum fluctuations. These results open prospects of forming pure liquid droplets of a quantum gas in free space characterized by a total absence of growth. 

\begin{acknowledgments}
We acknowledge insightful discussions with L. Santos and D. Petrov as well as with H.P B\"uchler, A. Pelster, G. Raithel and F. Ferlaino. We thank T. Maier for experimental assistance at early stages.  This work is supported by the German Research Foundation (DFG) within SFB/TRR21. 

\end{acknowledgments}

\bibliographystyle{apsrev4-1}

\bibliography{QuantumDroplets}
\newpage\null
\newpage

\section{Supplemental material}
\subsection{Experimental methods}
\begin{figure}[hbtp]
\includegraphics[width=.8\columnwidth]{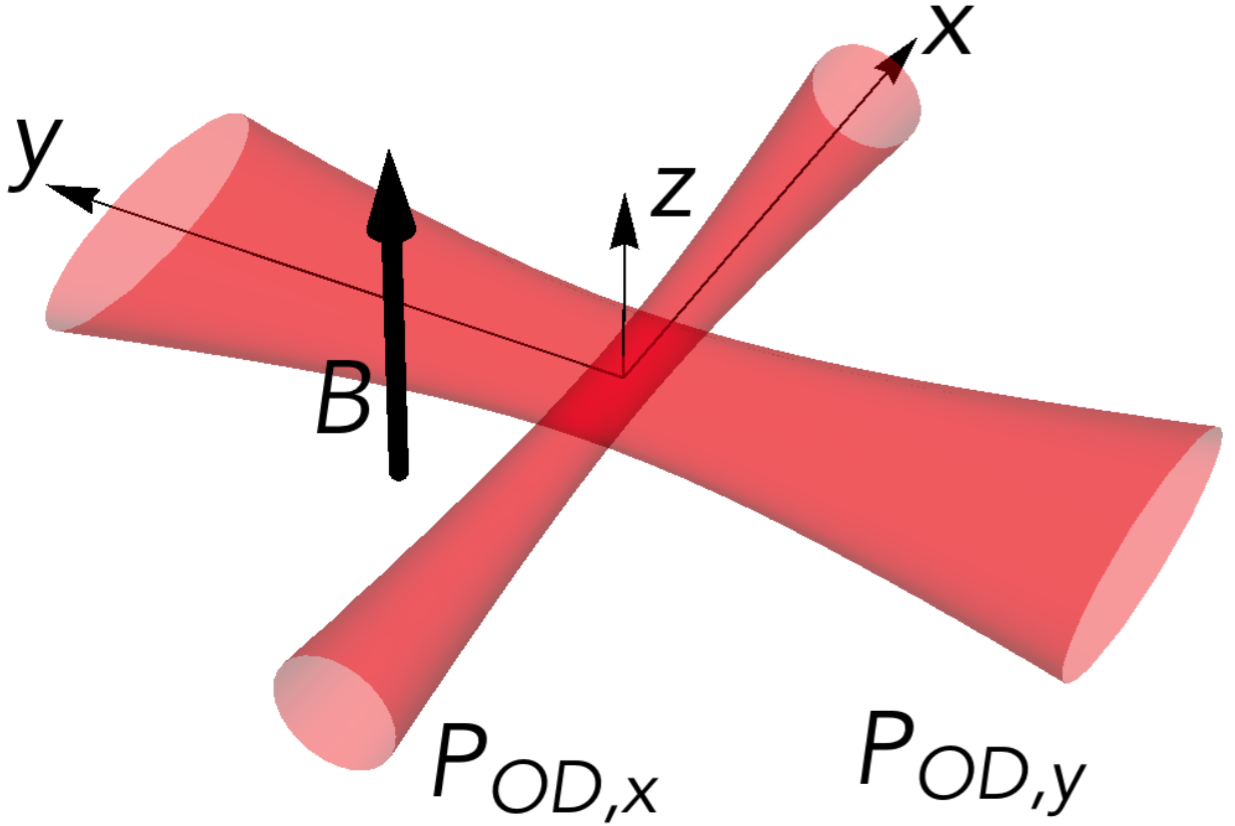}
\caption{Dipole traps and magnetic field in the experiment, the waveguide is obtained by turning off the laser propagating along $y$.}
\label{Fig:CrossedTrap}
\end{figure}
\begin{figure}[hbtp]
\includegraphics[width=\columnwidth]{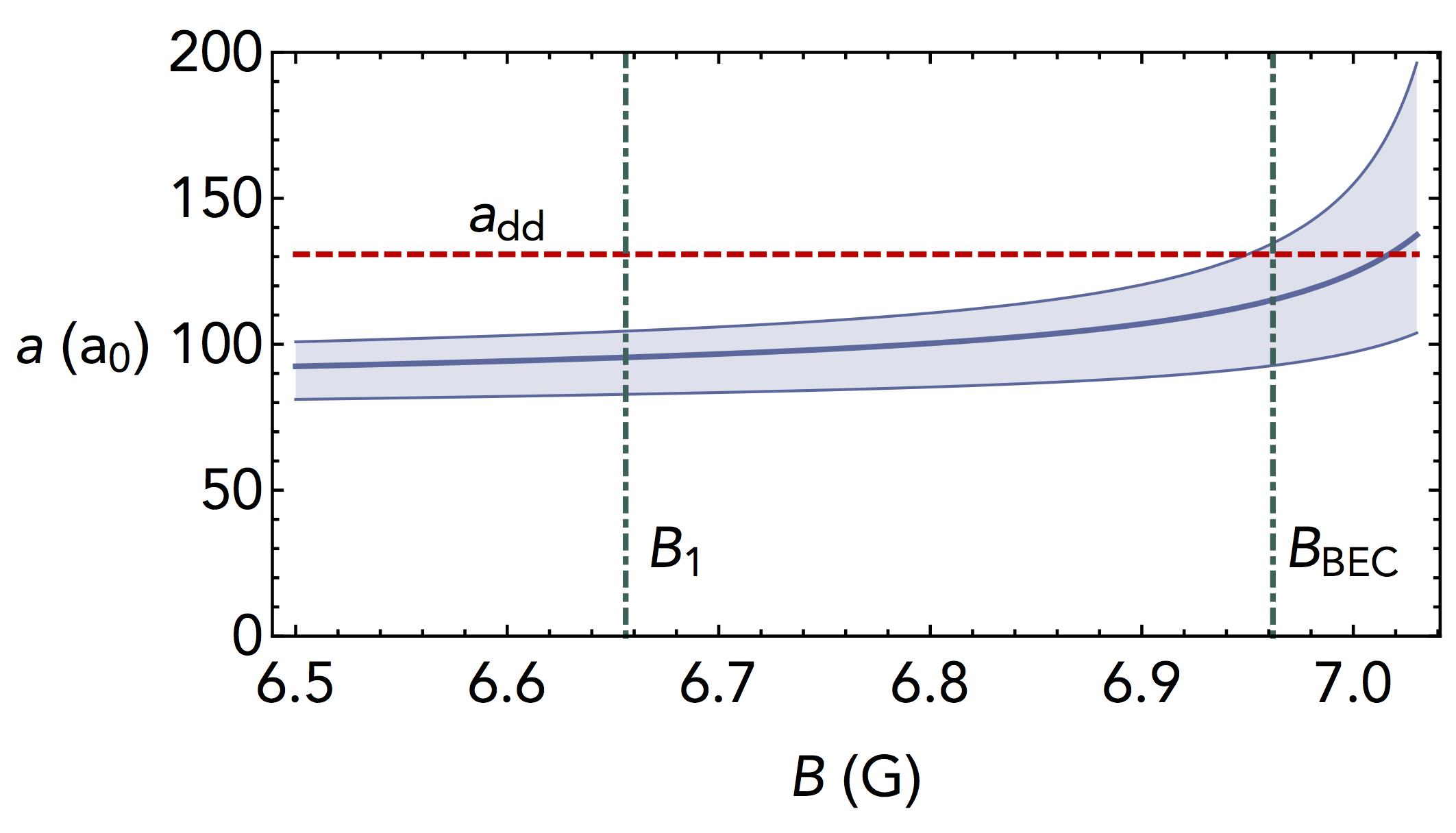}
\caption{Scattering length versus magnetic field with confidence interval, the dipole length is indicated in dashed red line. The fields at which we create the BEC $B_{\rm BEC}$ and observe the droplets $B_1$ are indicated as dot-dashed lines.}
\label{Fig:aofB}
\end{figure}
\subsubsection{Optical dipole trap and interaction control}

The main stages of our experiment have been presented in previous papers \cite{Maier:2015a,Maier:2015b,Kadau:2016}. We perform all the reported experiments using two dipole trap lasers, one propagating along the $x$ axis with waist $w=34(2)\,\mum$, the other one propagating along $y$ with asymmetric waists $w_x=103(5)\,\mum$ and $w_z=39(2)\,\mum$, fig.~\ref{Fig:CrossedTrap}. We create a waveguide by keeping only the first dipole laser. We also add a weak magnetic field gradient along the vertical $z$ axis to compensate partially gravity. This results in frequencies $\nu_y=123(10)\,\Hz$ and $\nu_z=100(10)\,\Hz$ in the waveguide. Residual magnetic field gradients pull the atoms away from the imaging region in this waveguide. To keep them within the field of view and measure their repulsion we kept the other dipole trap laser on at a very reduced power leading to an axial frequency $\nu_x=14.5(1)\,\Hz$. We place the atoms in a magnetic field oriented along $z$, at fields between $6.656(10)\,\g$ and $7.013(10)\,\g$. \dy~possesses many narrow Feshbach resonances that modify the scattering length and in ref.~\cite{Jachymski:2013} it was shown that the scattering length takes the form $a(B)=\abg\prod_i(1-\frac{\Delta_i}{B-B_{0,i}})$. We work close to a resonance that we calibrated in \cite{Kadau:2016}. Its position is $B_0=7.117(3)\,\g$ and its width is $\Delta=0.050(15)\,\g$, in addition, a relatively broad resonance exists at $B_0=5.1(1)\,\g$ with width $\Delta=0.1(1)\,\g$, which slightly pulls the scattering length down, other resonances are narrow or far enough away. Using the knowledge of the background scattering length reported in the main text $\abg=92(8)\,a_0$ we can back-out $a(B)$, plotted with error bars in fig.~\ref{Fig:aofB}. The error interval $\delta a$ on $a$ is given by a quadratic sum of all errors (on $\abg$, the positions $B_{0,i}$ and the widths $\Delta_i$),

\vspace{-.2cm}
\begin{scriptsize}
\be
\left(\frac{\delta a}{a}\right)^2=\left(\frac{\delta \abg}{\abg}\right)^2+\sum_i\left(\frac{\Delta_i}{\Delta_i+B-B_{0,i}}\right)^2\left(\left(\frac{\delta \Delta_i}{\Delta_i}\right)^2+\left(\frac{\delta B_{0,i}}{B_{0,i}}\right)^2\right).
\ee
\end{scriptsize}

Throughout the paper we present results from different ramping procedures for the magnetic field and trap powers that we present in the following. In figure 1 (a,b,c) of the main text we show droplets formed in a waveguide with no axial confinement. This is obtained starting in a crossed configuration creating an oblate trap cylindrically symmetric with aspect ratio $\nu_z/\nu_r\simeq2.9$, first by ramping down the magnetic field from $6.696(10)\,\g$ to $6.656(10)\,\g$ in $1\,\ms$, then turning off one dipole trap and increasing the waveguide laser power to obtain the above trapping frequencies, ramps are qualitatively represented in figure \ref{Fig:Ramp1} left panel. 
\subsubsection{Density measurement}

\begin{figure}[htbp]
\centerline{\includegraphics[width=0.5\columnwidth]{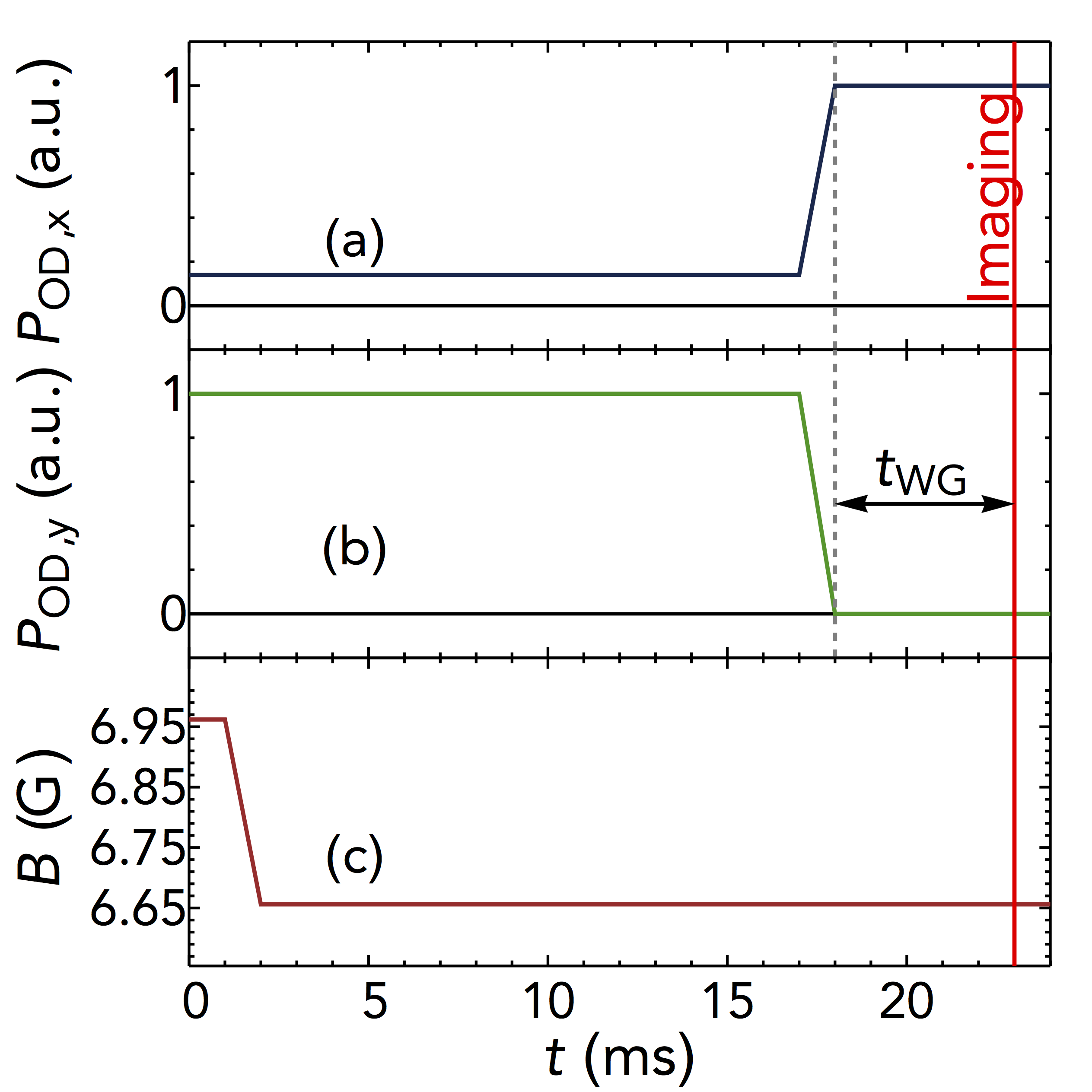}
\includegraphics[width=0.5\columnwidth]{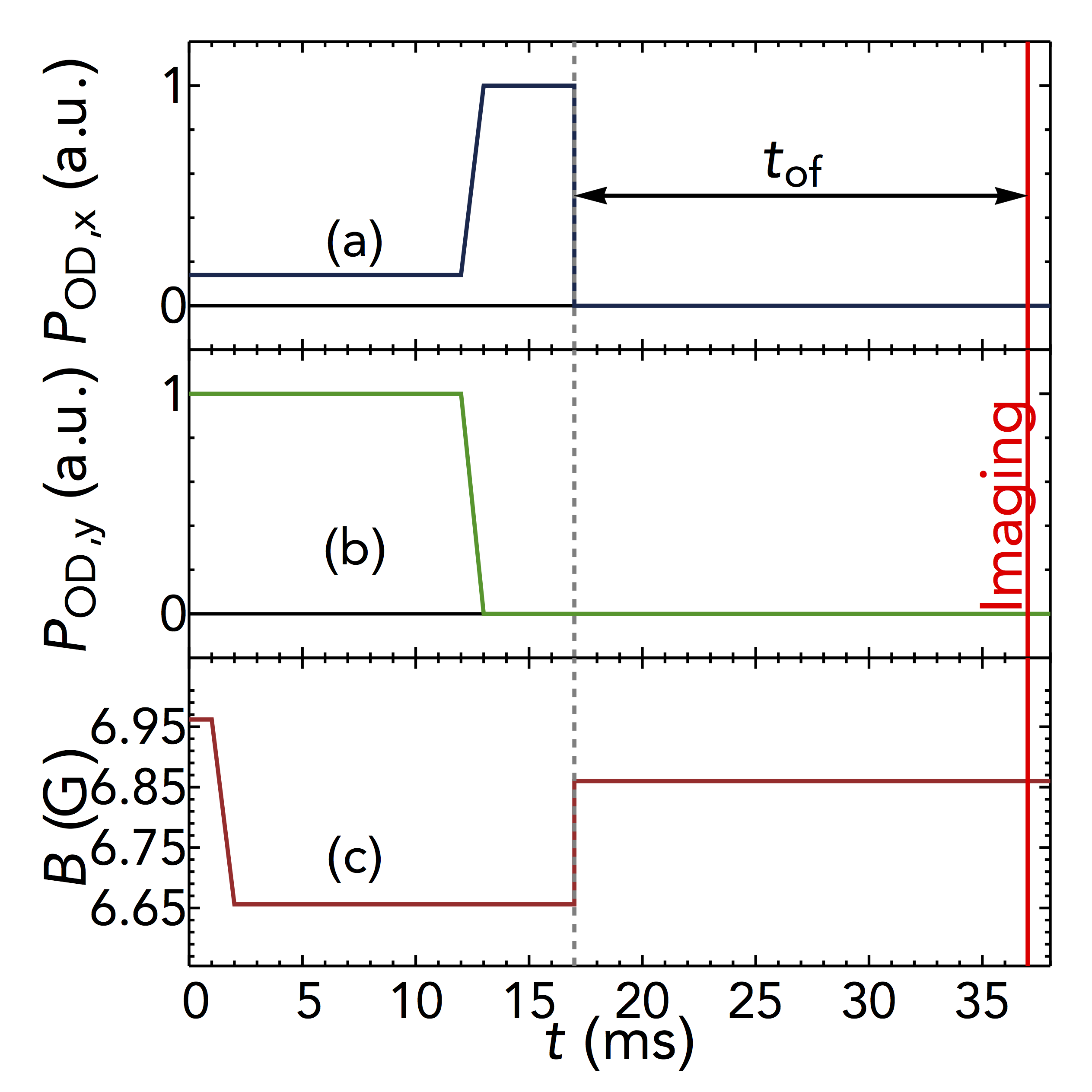}
}
\caption{Ramps of dipole trap powers (a,b) and magnetic field (c) performed for the data represented in the main text fig.~1(a-c), left panel and fig.~3 of main text, right panel. See also fig.~\ref{Fig:CrossedTrap}.}
\label{Fig:Ramp1}
\end{figure}

\begin{figure}[htbp]
\includegraphics[width=\columnwidth]{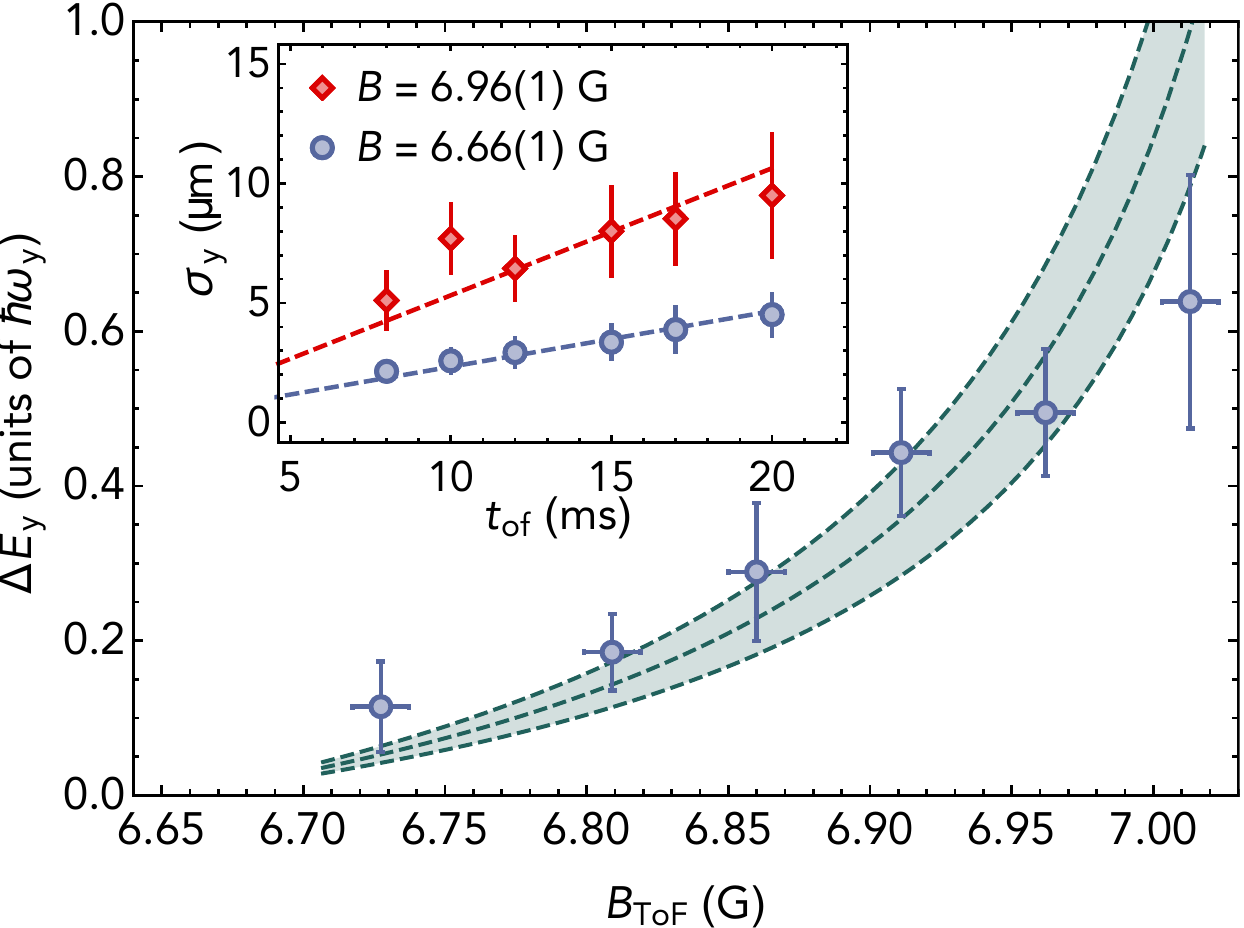}
\caption{Inset: gaussian width $\sigma_y$ in the $y$ direction as a function of time of flight $t_{\rm of}$ at $B=B_1$ (no quench, blue circles) and $B=6.96\,\g$ (red diamonds). By assuming a linear time dependence we evaluate the released energy $E_y=\frac12m\dot\sigma_y^2$ \cite{Holland:1997}. Main panel: Difference in released energy with respect to $B=B_1$, blue circles in units of the harmonic oscillator energy. Error bars represent the mean square residual. We compare this with an estimate of the change in released energy (eq.\ref{Eq:ReleasedEnergy}), from which we get $n_0=4.9(2.0)\times10^{20}\,{\rm m}^{-3}$. The shaded area represents this confidence interval.}
\label{Fig:ReleasedEnergy}
\end{figure}

In the main text we present a measurement of the density, which is obtained through the change in released energy due to a quench of the magnetic field at the beginning of time-of flight. The linear expansion of the droplets'size in time of flight from which we extract the released energy is represented for two different magnetic fields in the inset of figure \ref{Fig:ReleasedEnergy}. These data are obtained using the ramps shown in fig.~\ref{Fig:Ramp1} right panel, the fast quench in magnetic field is done in $50\,\mus$ simultaneously with release. 

\begin{figure}[hbtp]
\includegraphics[width=.85\columnwidth]{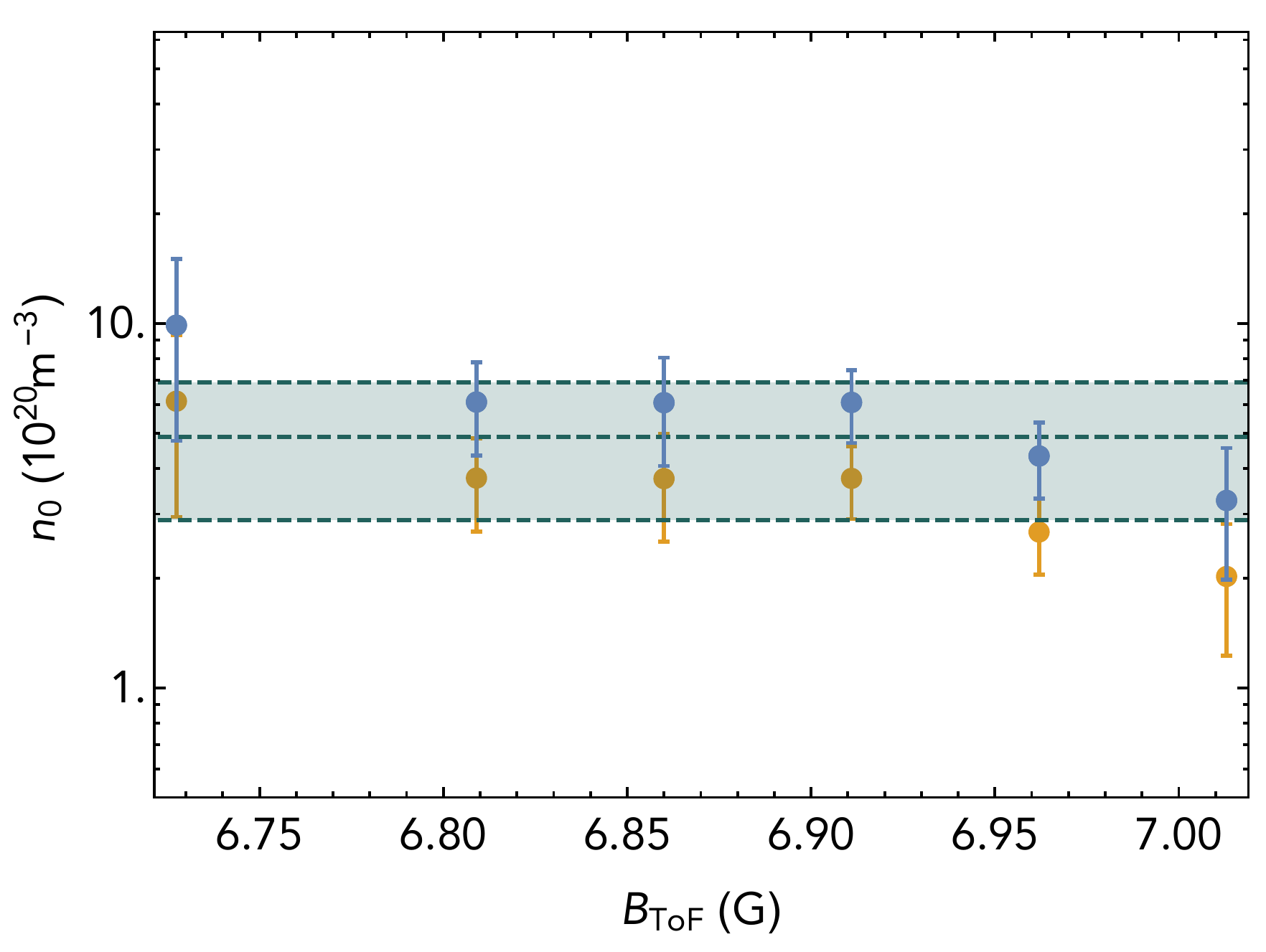}
\caption{Density $n_0$ extracted for each point of fig.\,\ref{Fig:ReleasedEnergy} using eq.~(\ref{Eq:density1}) for the blue points and (\ref{Eq:density2}) for the yellow ones. The weighted average and confidence interval is showed as dashed line and shaded area (using the blue points).}
\label{Fig:density}
\end{figure}

From these ramps we get the change in released energy at various magnetic fields, which is represented in figure \ref{Fig:ReleasedEnergy}. From these we extract the density using the following equation:
\be
\Delta E\simeq\frac1N\,\int d\vec r\;\frac{\Delta g}{2}  n^2=\frac{\Delta g}{2}\langle n\rangle\label{Eq:ReleasedEnergy}
\ee
resulting in
\be
n_0=4\sqrt2\,\Delta E/\Delta g\label{Eq:density1}
\ee
for the gaussian ansatz, and
\be
n_0=7\,\Delta E/2\,\Delta g\label{Eq:density2}
\ee
for an inverted parabola thus from each point we extract a density and an uncertainty $\delta n_0$, with $\left(\frac{\delta n_0}{n_0}\right)^2=\left(\frac{\delta \Delta E}{\Delta E}\right)^2+\left(\frac{\delta \Delta g}{ \Delta g}\right)^2$. These points are represented in fig.~\ref{Fig:density} (blue: gaussian ansatz, yellow: inverted-parabola ansatz), we then take the error-weighted mean to obtain $n_0$, and show the confidence interval in fig.~\ref{Fig:density}. We add to this error a systematic error accounting for the fact that our released-energy model is a simplification of the full free-space dynamics. Note that in fig.~\ref{Fig:ReleasedEnergy} the only apparent error on the vertical axis is due to the uncertainty on $\Delta E$ but not the one on $\Delta g$, this one can only be apparent in the vertical axis of fig.~\ref{Fig:density}. For this reason the confidence interval shown is for a fixed $\Delta g (B)$.

\subsubsection{Lifetime measurements}

\begin{figure}[htbp]
\includegraphics[width=\columnwidth]{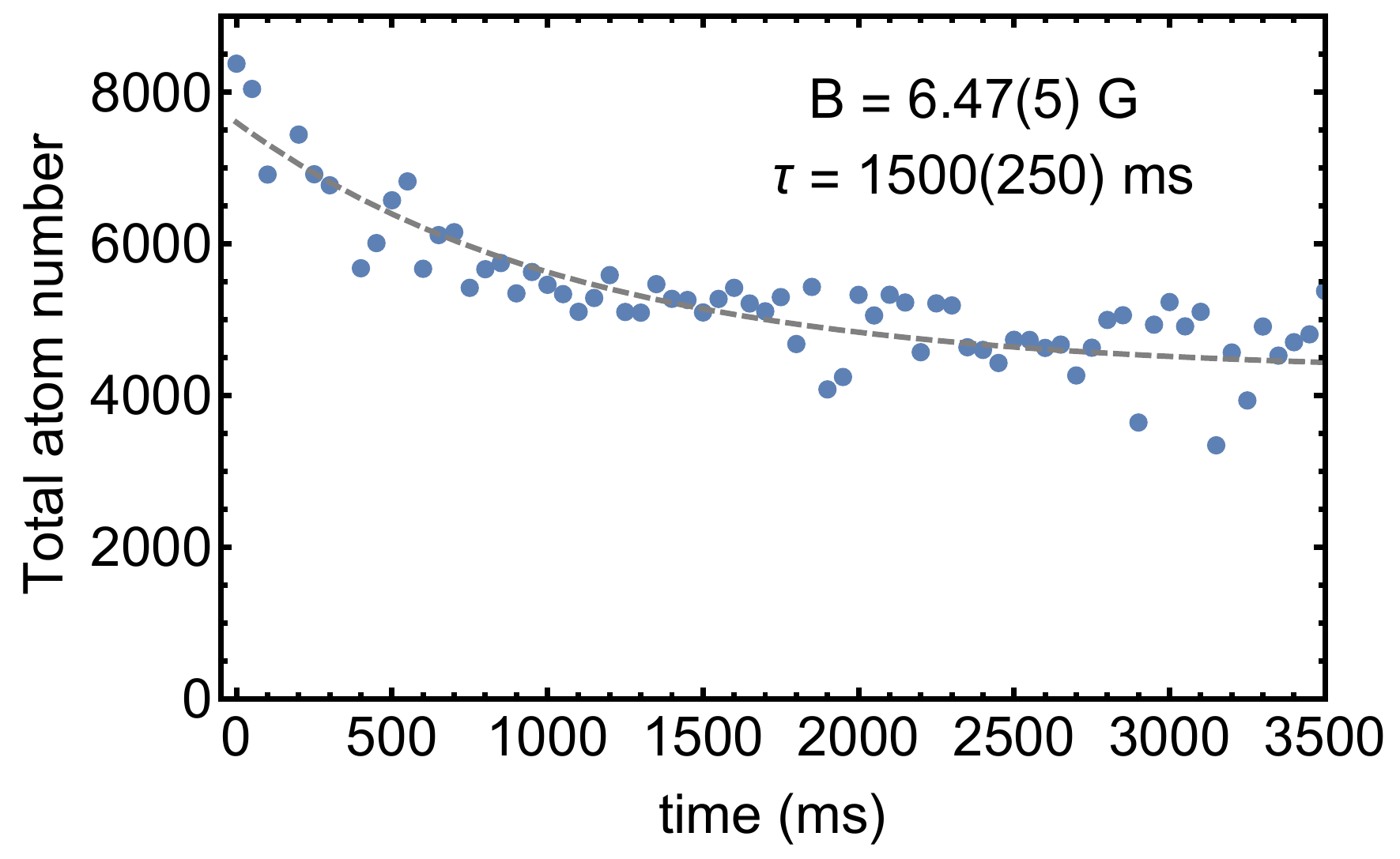}
\includegraphics[width=\columnwidth]{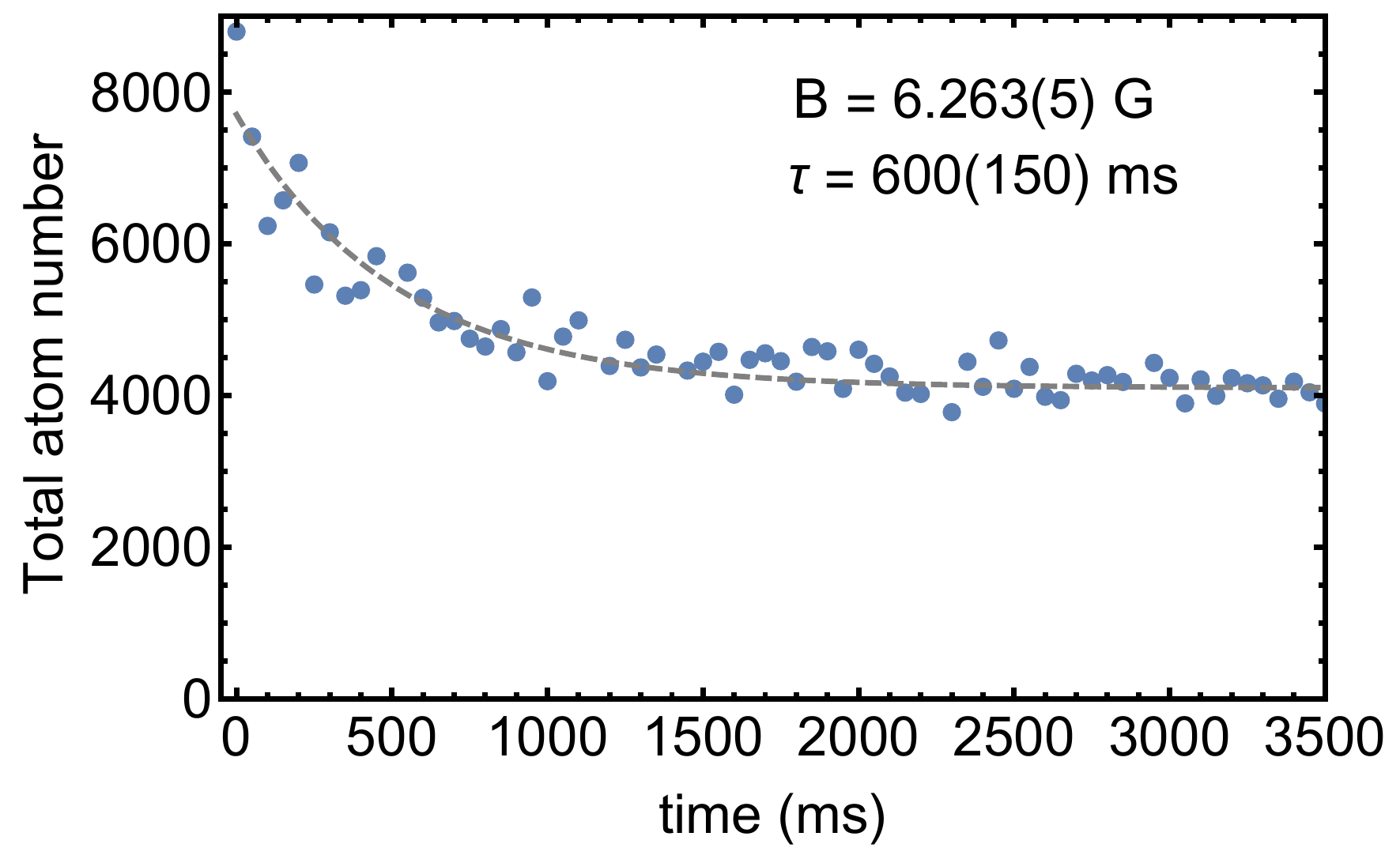}
\includegraphics[width=\columnwidth]{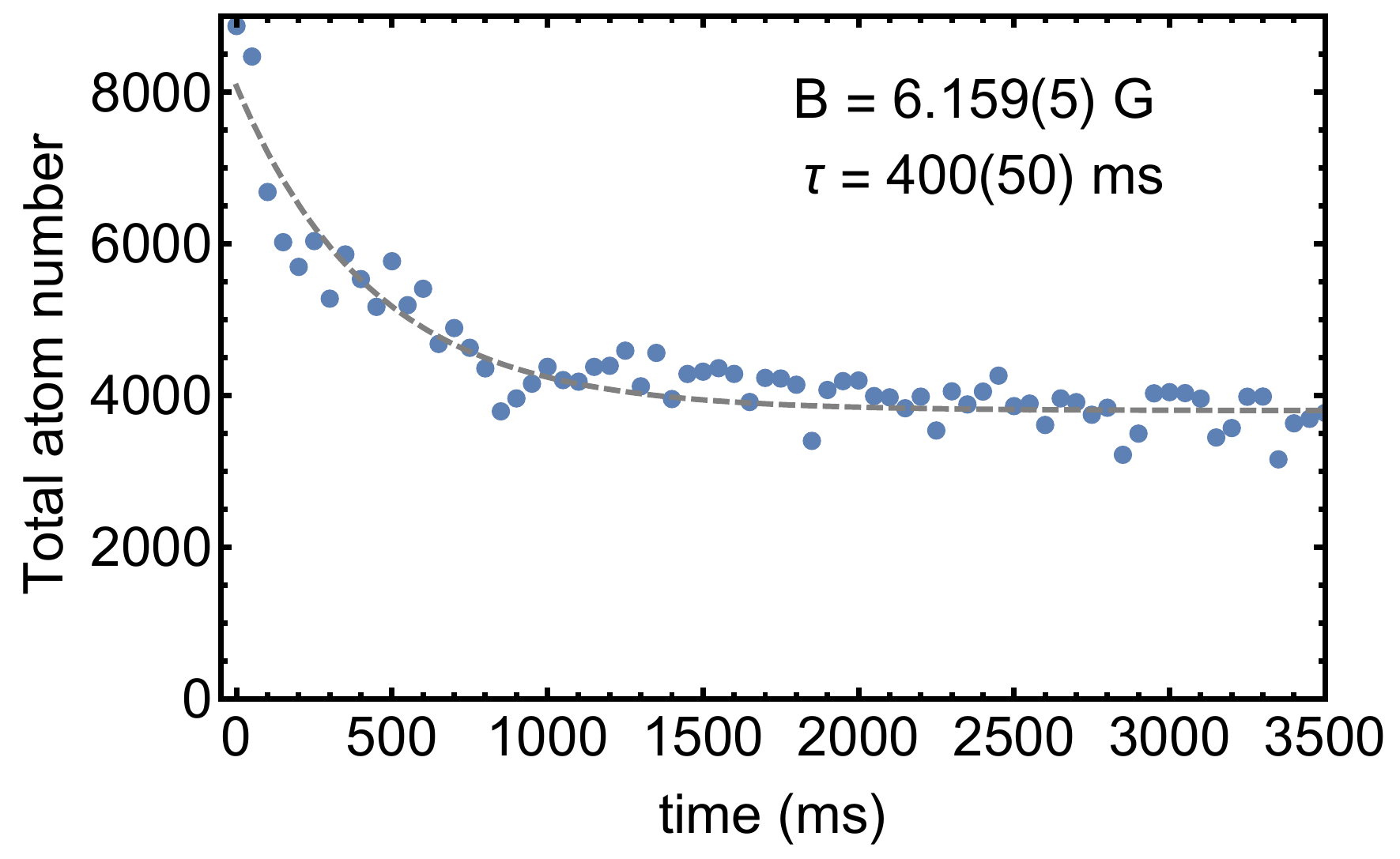}
\caption{Atom number lifetime measurements. Atom number in the cigar-shaped trap as a function of time after a field quench through the instability at three different magnetic fields. The existence of two timescales is evident on the top trace. From the fast initial dynamics we extract the droplets lifetime $\tau$.}
\label{Fig:Lifetimes}
\end{figure}

We report on the measurement of the scaling of the lifetime of the droplets versus scattering length or equivalently magnetic field. These measurements are all performed in the cigar-shaped trap after a sudden quench of the magnetic field through the instability (as opposed to the ones reported in ref.\,\cite{Kadau:2016}), where we record the total atom number in our atomic clouds which contain several droplets. Experimentally we observe two time scales, a fast one which we call $\tau$ (typically a few hundreds of $\ms$ to a few seconds) and a very slow one of several seconds, which is exemplified in fig.\,\ref{Fig:Lifetimes}. We associate the fast one to the lifetime of the droplets, before a remnant cloud is left containing between 4000 atoms which is too dilute to form droplets and thus decays much more slowly. We do in fact observe droplets only during the initial fast decay. The lifetime of the droplets is then extracted by fitting an exponential decay to the initial atom loss. By allowing the long-term atom number to change between 3750 and 4250 we extract an uncertainty on $\tau$.

%\begin{figure}[htbp]
%\includegraphics[width=\columnwidth]{Ramp2.pdf}
%\caption{Ramps of dipole trap powers and magnetic field performed for the data represented in the main text fig.~3. See also fig.~\ref{Fig:CrossedTrap}.}
%\label{Fig:Ramp3}
%\end{figure}

%\newpage\null
%\newpage\null
\newpage

\subsection{Models}
\subsubsection{Droplet elongation}
\normalsize
We observe a spacing of $d=2.5(5)\,\mum$ between droplets at equilibrium in a trap with frequency $\nu_x=14.5\,\Hz$. As stated in the main text this distance cannot be accounted for by assuming two point-like droplets repelling each other. We thus conclude that the distance is reduced by the finite extent of the droplets in the z direction. To quantitatively account for this effect we calculate the dipole-dipole energy $E_{\rm dd}(d)$ between two droplets at a distance $d$ in the gaussian ansatz: assuming a density distribution $n(\vec x)=\frac{N}{(2\pi)^{3/2}\sigma_r^2\sigma_z}\exp\left(-\frac{r^2}{2\sigma_r^2}-\frac{z^2}{2\sigma_z^2}\right)$, with $N$ the number of atoms per droplet. $E_{\rm dd}(d)$ is then given by\\

\vspace{-.9cm}
\be
\begin{split}
E_{\rm dd}(\vec d\,)=\int d\vec x_1\,n(\vec x_1-\vec d/2)\int d\vec x_2\Big[ \\
\left.n(\vec x_2+\vec d/2)V_{\rm dd}(\vec x_1-\vec x_2)\right].
\end{split}
\ee
This repulsion is counteracted by the trap energy which reads
\begin{small}
\be
E_{\rm trap}(d\,)=\frac{\mu_R\omega^2d^2}{2}
\ee
\end{small}
with $\mu_R=Nm/2$ the reduced mass of two droplets. Using similar calculations as in \cite{Fattori:2008} we get:\\

\begin{figure}[htbp]
\includegraphics[width=\columnwidth]{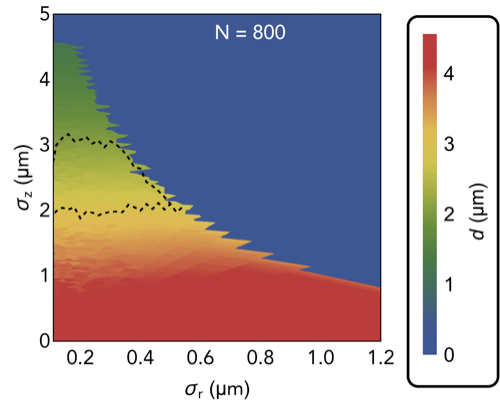}
\caption{Distance of the local minimum between two neighbouring droplets in the $\sigma_r,\,\sigma_z$ plane, this local minimum exists only in a certain range of parameters. One sees that the position of the minimum when it exists depends only on $\sigma_z$. The area surrounded by a dashed line shows the $\sigma_z$, $\sigma_r$ values for which $2\,\mum\leqslant d\leqslant 3\,\mum$.}
\label{Fig:dVsSigmas}
\end{figure}

\begin{figure}[htbp]
\includegraphics[width=\columnwidth]{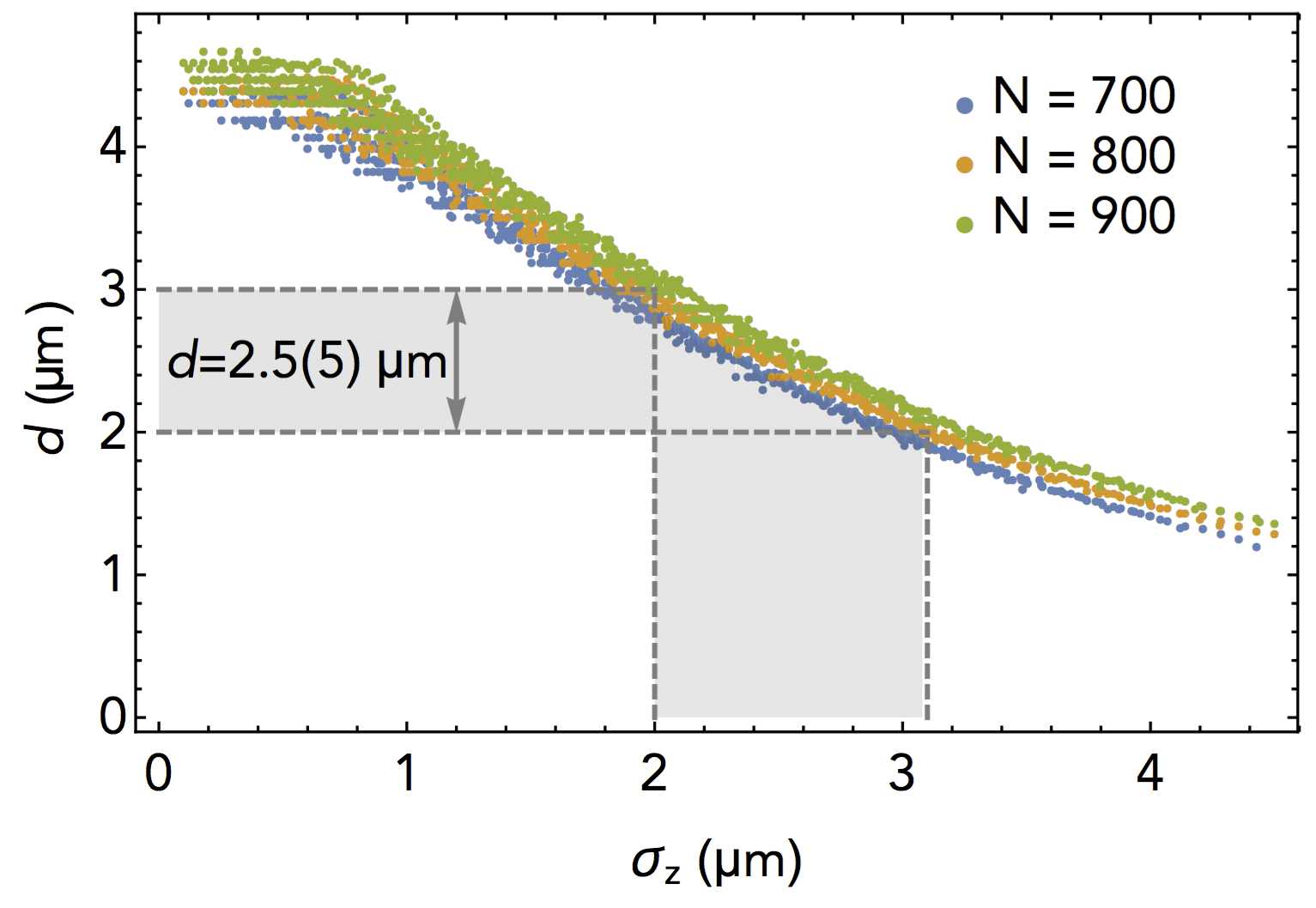}
\caption{Distance of the local minimum between two neighbouring droplets as a function of $\sigma_z$ when a local minimum exists. Using our measure of $d=2.5(5)\,\mum$ (grey horizontal area) we obtain bounds on $\sigma_z$ (grey vertical area). The three different colors are for three different atom numbers per droplet (700, 800, 900), one sees only a weak dependence in $N$.}
\label{Fig:dVsSigmaZ}
\end{figure}

\vspace{-.9cm}
\be
&&E_{\rm dd}(d)\,=\,\frac{\mu_0\mu^2N^2}{3(2\pi \sigma_r)^3}\,I(\lambda,\kappa)\\
&&\kappa=\frac{\sigma_r}{\sigma_z},\;\lambda=\frac{d}{\sigma_r}
\ee
\be
\begin{split}
I(\lambda,\kappa)=4\pi\int_0^{\infty}dv\,\int_0^1du\,\Big[v^2(1-3u^2)\times\\
\left.J_0(\lambda v\sqrt{1-u^2})\,\exp\left(-v^2(1-u^2(1-\kappa^{-2}))\right)\right]
\end{split}
\ee
with $J_0$ the Bessel function of the first kind. We then look for the position of the minimum in energy as a function of $d$ which gives us the distance at which two neighbouring droplets equilibrate. Note that this is only a local minimum, since for our parameter range two neighbouring droplets always attract each other at very short distances. In fig.~\ref{Fig:dVsSigmas}, we represent the distance obtained as a function of $\sigma_r$ and $\sigma_z$ for 800 atom in the droplets. A local minimum exists only for low enough sizes. In the absence of a local minimum droplets should always attract each other. When a local minimum exists, its position depends only on $\sigma_z$ as can be seen in fig.~\ref{Fig:dVsSigmas}, we thus obtain only an upper bound on $\sigma_r$, $\sigma_r\leqslant500\,\nm$. We plot in fig.~\ref{Fig:dVsSigmaZ} the dependence of $d$ on $\sigma_z$ when a local minimum exists. From this we see that our experimental measure on $d=2.5(5)\,\mum$ leads to $\sigma_z=2.5(5)\,\mum$.\\


\subsubsection{Droplet stability}

We discuss in the main text the stability of the droplets against mean-field collapse. For this purpose we use the Thomas-Fermi approximation which neglects the kinetic energy (see discussion below). In this approximation the problem of calculating the mean-field chemical potential of a inhomogeneous dipolar BEC has been treated, see for instance eq.~(4.5) of ref.~\cite{Lahaye:2009} or eq.~(49) of ref.~\cite{Lima:2012}. This mean-field chemical potential reads 
$\mu(\vec x)=gn(\vec x)+\Phi_{\rm dd}(\vec x)+V_{\rm ext}(\vec x)$ with $g=4\pi\hbar^2 a/m$. The DDI contribution is
\be
\Phi_{\rm dd}(\vec x)=\int d\vec {x'}V_{\rm dd}(\vec x-\vec{x'})n(\vec{x'}).
\ee
We consider now the central density $n_0$. Using the gaussian ansatz $n(\vec x)=n_0\,\exp\left(-\frac{r^2}{2\sigma_r^2}-\frac{z^2}{2\sigma_z^2}\right)$ \cite{Lahaye:2009} or an inverted parabola $n(\vec x)=n_0\,\left(1-\frac{r^2}{R_r^2}-\frac{z^2}{R_z^2}\right)$ \cite{Odell:2004} one obtains the well-known result
\be
\mu_{\rm mf,dd}&=&gn_0\,\edd\, f(\kappa),
\ee
with
\be
f(\kappa)&=&\frac{1+2\kappa^2}{1-\kappa^2}-\frac{3\kappa^2	\,{\rm arctanh}(\sqrt{1-\kappa^2})}{(1-\kappa^2)^{3/2}}.
\ee
Calculating then the compressibility at center $\frac{\partial \mu}{\partial n_0}$ one obtains that for our values of $\kappa$ the droplets should be unstable. However the mean-field approximation has to be corrected by the beyond-mean-field term originating from quantum fluctuations $\mu_{\rm bmf}\simeq\frac{32\,gn}{3\sqrt\pi}\sqrt{na^3}(1+\frac32\edd^2)$, by doing so we get eq.~(1) from the main text.

\subsubsection{Droplet lifetime}
The droplets are dense and as such are prone to three-body recombination losses. The loss equation in this case is given by 
\be
\frac{dN}{dt}=-L_3\langle n^2\rangle N,\label{Eq:3bLosses}
\ee
With $N$ the total atom number. A peculiar feature of our droplets is that the density does not depend on the total atom number $N$ as is visible in eq.\,(2) from main text. Thus the solution to eq.\,(\ref{Eq:3bLosses}) is a simple exponential decay with characteristic time 
\be
\tau=1/L_3\langle n^2\rangle
\ee
this characteristic time depends on the precise value of $L_3$ as well as on the exact density distribution $n(\vec r)$ in a droplet. One however does not need that knowledge to observe the scaling of the density (or $\tau$) with scattering length as noted in main text. Using the relation $\langle n^2\rangle\propto n_0^2$ the ratio of the lifetime at two different scattering lengths $a_{\rm i}$ and $a_{\rm f}$ is given by $\tau_{\rm f}/\tau_{\rm i}=\frac{n_{0, \rm i}^2}{n_{0, \rm f}^2}$. This result does make the assumption that $L_3$ is independent on $a$. In our situation of \textsuperscript{164}Dy close to the background scattering length, $a$ is not the dominant length scale of the interactions since the dipole length and the Van der Waals length in the units of Gao \cite{Gao:2008} (see supplemental material of \cite{Maier:2015a}) are respectively $392\,a_0$ and $154\,a_0$. As a result one does not expect three-body recombination to depend on $a$ \cite{Wang:2014,Shotan:2014}, we have verified this by three-body recombination measurements in thermal gases which will be presented in future publications. In conclusion, using eq.\,(2) from main text one can calculate $\tau_{\rm f}/\tau_{\rm i}$ without difficulty, for a fixed $\kappa$ it depends only on two parameters which we chose to be 
\be\varepsilon_{\rm dd, i}=\frac{\add}{a_{\rm i}}\ee
and 
\be a_{\rm f/i}=\frac{a_{\rm f}}{a_{\rm i}}
\ee
then one has 
\be
\frac{\tau_{\rm f}}{\tau_{\rm i}}=\left(a_{\rm f/i}\right)^6\left(\frac{\varepsilon_{\rm dd,i}\, f_{\rm dip}(\kappa)-1}{\frac{\varepsilon_{\rm dd,i}}{a_{\rm f/i}}\, f_{\rm dip}(\kappa)-1}
\frac{1+\frac32\,\left(\frac{\varepsilon_{\rm dd,i}}{a_{\rm f/i}}\right)^2}{1+\frac32\,\varepsilon_{\rm dd,i}^2}\right)^4
\ee
assuming three-body repulsion as a stabilizing mechanism this becomes
\be
\frac{\tau_{\rm f}}{\tau_{\rm i}}=\left(a_{\rm f/i}\right)^{-2}\left(\frac{\varepsilon_{\rm dd,i}\, f_{\rm dip}(\kappa)-1}{\frac{\varepsilon_{\rm dd,i}}{a_{\rm f/i}}\, f_{\rm dip}(\kappa)-1}\right)^2
\ee
These two expressions are represented in fig.\,(3) of main text as filled blue and hatched green, respectively. They do depend on $\kappa$, though weakly, which is why we plot the whole range obtained when varying $\kappa$ between $0$ and $1/5$ the experimental lower bound. Even with this uncertainty we eliminate three-body repulsion and validate quantum fluctuations as a stabilizing mechanism.
\bibliographystyle{apsrev4-1}

\end{document}